# Network pharmacology on the mechanism of Yi Qi Tong Qiao Pill inhibiting allergic rhinitis

Boyang Wang[a], DingFan Zhang[a], Tingyu Zhang[a], Chayanis Sutcharitchan[a], Jianlin Hua[b], Dongfang Hua[b,*], Bo Zhang[c,*], Shao Li[a,*]

[a]Institute for TCM-X, MOE Key Laboratory of Bioinformatics, Bioinformatics Division, BNRist, Department of Automation, Tsinghua University, 100084 Beijing, China

[b]Tianjin Oriental HuaKang Pharmaceutical Technology Development Co., Ltd., 300457 Tianjin, China

[c]TCM Network Pharmacology Department, Tianjin Key Laboratory of Early Druggability Evaluation of Innovative Drugs, Tianjin International Joint Academy of Biomedicine, 300457 Tianjin, China

*Correspondence should be addressed to:

Prof. Shao Li,

FIT 1-115, Tsinghua University, 10084, Beijing

Email: shaoli@mail.tsinghua.edu.cn

Dr. Bo Zhang,

Tianjin International Joint Academy of Biomedicine, 300457, Tianjin

Email: zhangbo@tjab.org

Dr. Dongfang Hua,

Tianjin Oriental HuaKang Pharmaceutical Technology Development Co., Ltd., 300457, Tianjin

Email: 13903592082@163.com

Abstract:

Objective: The purpose of this study is to reveal the mechanism of action of Yi Qi Tong Qiao Pill (YQTQP) in the treatment of allergic rhinitis (AR), as well as establish a paradigm for the researches on traditional Chinese medicine (TCM) from systematic perspective.

Methods: Based on the data collected from TCM-related and disease-related databases, target profiles of compounds in YQTQP were calculated through network-based algorithms and holistic targets of TQTQP was constructed. Network target analysis was performed to explore the potential mechanisms of YQTQP in the treatment of AR and the mechanisms were classified into different modules according to their biological functions. Besides, animal and clinical experiments were conducted to validate our findings inferred from Network target analysis.

Results: Network target analysis showed that YQTQP targeted 12 main pathways or biological processes related to AR, represented by those related to IL-4, IFN-γ, TNF-α and IL-13. These results could be classified into 3 biological modules, including regulation of immune and inflammation, epithelial barrier disorder and cell adhesion. Finally, a series of experiments composed of animal and clinical experiments, proved our findings and confirmed that YQTQP could improve related symptoms of AR, like permeability of nasal mucosa epithelium.

Conclusion: A combination of Network target analysis and the experimental validation indicated that YQTQP was effective in the treatment of AR and might provide a new insight on revealing the mechanism of TCM against diseases.

# Introduction

Allergic rhinitis (AR) is a chronic inflammatory disease of the upper respiratory tract, characterized by symptoms such as sneezing, clear nasal discharge, nasal itching, and nasal congestion[1, 2]. Its pathogenesis involves immediate hypersensitivity mediated by IgE-bound mast cells, as well as a late-phase reaction characterized by the recruitment of eosinophils, basophils, and T cells producing interleukin (IL)-4 and IL-5[3]. Although not a serious illness, AR has a great impact on patients' quality of life and poses as a major risk factor for poor asthma control [2]. Various form of $H_1$-antihistamines and corticosteroids, including oral, intranasal, ocular, are commonly used to treat AR in clinical practice. However, these pharmacological treatments suffer certain drawbacks such as side effects, which limit the use among some patient groups, and the inability to eradicate the condition, thus require long-term use to control the symptoms [1]. Consequently, more effective alternatives still need to be sought.

Yi Qi Tong Qiao Pill (YQTQP) is an empirical TCM formula for treating AR. It is composed of 14 TCM herbs, including Astragali Radix (Huangqi), Saposhnikoviae Radix (Fangfeng), Magnoliae Flos (Xinyi), Angelicae Dahuricae Radix (Baizhi), and Atractylodis Macrocephalae Rhizoma (Baizhu). On the basis of TCM theories, the symptoms of AR such as nasal congestion are caused by lung-spleen qi deficiency, weakened defensive qi, and external wind invasion. YQTQP can reinforce qi, consolidate the superficial resistance, and disperse the wind, therefore, it is commonly used to treat conditions equivalent to AR in TCM practice. Previous studies have shown that YQTQP can effectively improve the clinical symptoms of AR, but there is still a lack of research on its micro-molecular mechanisms and biomarkers[4-6].

Network pharmacology of traditional Chinese medicine approaches molecular relationship between drugs and diseases from holistic perspective[7-9]. Through the concept of "Network target", it provides a systemic perspective on the mechanism of drug regulation on biological systems[10-12]. Based on this theory, the UNIQ (Using Network target for Intelligent and Quantitative analysis on drug actions) system[13] has been developed as an intelligent and quantitative analysis system for network pharmacology. This system has been successfully applied to the systematic study of complex diseases, like inflammation-induced tumorigenesis[14] and hepatocellular carcinoma[15], as well as various complex formulas, including WFC[16] and MLD[17]. Since 2015, we conducted a series of studies for the research and development of YQTQP, as well as uncovering the mechanism of action of YQTQP in the treatment of AR through UNIQ system and experiments validation composed of animal and clinical experiments.

# 1 Materials and methods

### 1.1 Data collection

Data on chemical composition of each herb in YQTQP was collected from HERB database[18]. The collected raw data was deduplicated to obtain overall chemical components contained in YQTQP, and corresponding information of each component was subsequently obtained from PubChem database. Data on AR was collected from Comparative Toxicogenomics Database (CTD)[19] and MalaCards[20].

### 1.2 Target prediction and validation

We performed a chemical similarity and network-based drug target prediction algorithm, DrugCIPHER, to predict the genome-wide targets of each chemical component in YQTQP. The top100 of the predicted targets was defined as the target profile of the compound. The co-occurrence relationships of each chemical component and its target profile were uncovered through the abstracts of literatures in the PubMed database and the drug-target relationships recorded in the PubChem database were also collated with the co-occurrence results presented as the literature coverage of the predicted targets. The existing literature coverage was applied to evaluate the accuracy of the target prediction for each compound. The accuracy was calculated by:

$$\frac{\text{(the number of the intersection of the predicted targets and reported biomolecules)}}{\text{the number of predicted targets}} \times 100\%$$

**1.3 Network target analysis and multi-level network construction**

Based on statistical model previously established by Liang et al.[21], the holistic targets of YQTQP were identified. The model based on the assumption that "the more targets that appear in the main components of YQTQP, the more likely they are to be the key targets". Subsequently, KEGG and GO gene set enrichment analysis was performed to identify key biological pathways and processes of YQTQP in the treatment of AR. The potential biological pathways and processes associated with the pathogenesis of AR were selected from the enriched pathways. The selected pathways were grouped in to key target biological modules. Key molecules associated with the pathways were imported into STRING database to obtain PPI network for each module. The networks of each module were then drawn in Cytoscape, based on their connectivity. Three TCM herbs modules of related networks were constructed according to the compatibility relationship of YQTQP, and the enrichment of herb's potential targets contained in the key modules of YQTQP were calculated. Finally, the herbs were connected based on the modules the related targets they were enriched in.

**1.4 Animal study**

Fifty Kunming mice, half male and half female, weighing between 18-22g, were randomly divided into five groups based on gender and body weight balance: high, medium, and low dose of the test drug (27, 13.5, and 6.75 g of crude drug per kg), saline, and loratadine (2mg/kg). Each mouse was sensitized with 0.3 ml of Freund's incomplete adjuvant (10 mg/ml) and 0.5 ml of intraperitoneal (i.p) solution containing 50 μg/ml of smallpox pollen protein and 0.3 ml of subcutaneous injection of Freund's incomplete adjuvant. After sensitization, the mice were orally administered with the drug solution or saline once a day for 14 days.

A total of 50 rats with a body weight of 150±10g were randomly divided into five groups: high, medium, and low doses of the test drug (22.5, 11.25, 5.625 g crude drug/kg), chlorpheniramine maleate (2mg/kg), and physiological saline. The drug solution or physiological saline was administered orally once a day for 7 consecutive days.

60 minutes after the last administration, i.v 1% Evans blue normal saline solution 0.1ml/10g body weight, and immediately drip 20μl of 0.1% bovine serum albumin aqueous solution into each nostril. After 20 minutes, the mice were euthanized, and the nasal cavity was cut along the line of the inner canthus of both eyes, the blood was blotted with filter paper, the skin was peeled off, the turbinates and mucous membranes were cut into pieces, immersed in 3ml of 70% acetone saline for 24 hours, centrifuged at 2500 rpm for 10 minutes, and the supernatant was taken and measured for absorbance at 600 nm wavelength to detect the allergic effect on the nasal mucosa of mice.

60 minutes after the last administration, i.v 1% Evans blue normal saline solution 0.1ml/10g body weight, and immediately i.p trichosanthin aqueous solution (20ug/ml) 0.3ml/rat. After 20 minutes, the mice were euthanized, the peritoneal cavity was washed with normal saline, the washing solution was centrifuged at 2500 rpm for 10 minutes, and the supernatant was separated to measure the absorbance at 600 nm to detect the effect on the permeability of the peritoneal vessels of the trichosanthin-challenged mice

One hour after the last administration, 0.1ml of histamine phosphate (0.5mg/ml) was intradermally injected into the area of about 4cm2 of hair removal on the back, and 0.1ml of i.v. 1% Evans blue normal saline solution per 10g body weight was injected, and the head was decapitated after 20min. The colored part of the back skin was peeled off, cut it into pieces and soaked in acetone saline (7:3) solution for 24 hours, centrifuged at 2500 rpm for 10 minutes. The supernatant was taken and the absorbance was measured at 600nm to detect the permeability of rat skin capillaries induced by histamine sexual influence.

## 1.5 Clinical study

A randomized double-blind, high-low dose parallel control trial was carried out to assess the clinical efficacy of YQTQP for AR. Four-hundred eighty subjects were enrolled in the study. The subjects were patients aged 18-60 years old, diagnosed with seasonal AR, with a course of disease ≥ 2 years, met the TCM syndrome differentiation criteria of lung-qi deficiency, and voluntarily signed the informed consent form.

The subjects were randomly divided into high-dose and low-dose groups, which included 360 and 120 subjects, respectively. High-dose groups administered YQTQP equivalent to 4 g/10 pills of medicinal materials, whereas low-dose groups administered mock-up YQTQP equivalent to 0.4 g/10 pills of medicinal materials. All subjects administered 20 pills per time, 3 times per day, 0.5 hr after meal with warm water for 14 consecutive days.

Before and during the treatment, overall symptoms and quality of life were documented by each subject on a daily basis. Evaluation of primary efficacy indicator, total nasal syndrome score (TNSS), and secondary efficacy indicators, including improvement of overall nasal and ocular symptoms, TCM syndrome score, single symptom disappearance rate, and quality of life, were made by medical personnel before the start of the treatment as baseline, followed by assessment at day 7 and 14 of the treatment. The subjects whose clinical symptoms disappeared after the treatment was followed-up by telephone interviews 7 days after the last dose of the treatment.

**1.6 Statistic analysis method in experiments**

All statistical tests were two-sided, and P value less than 0.05 was considered statistically significant (unless otherwise stated). Descriptive statistics for continuous variables were presented as mean, median, standard deviation, maximum, minimum, and the 25th and 75th percentiles. Categorical or ordinal variables were described using frequency and frequency tables.

## 2 Results

**2.1 Target prediction and validation**

A total of 1839 chemical components in YQTQP were obtained from HERB

database. The genome-wide target profiles of each component was calculated by DrugCIPHER[22]. The literature coverage rate of predicted targets of major chemical components were 76-95% (Figure 1A). Taking Daidzein as an example, 93% of its target profile were covered by literature and databases, while the remaining 7% were in the same biological network constructed by the reported biomolecules (Figure 1C). These results indicated that the target profile predicted by DrugCIPHER had strong reliability and was able to comprehensively depict the mechanism and synergistic effects of the ingredients in YQTQP (Figure 1).

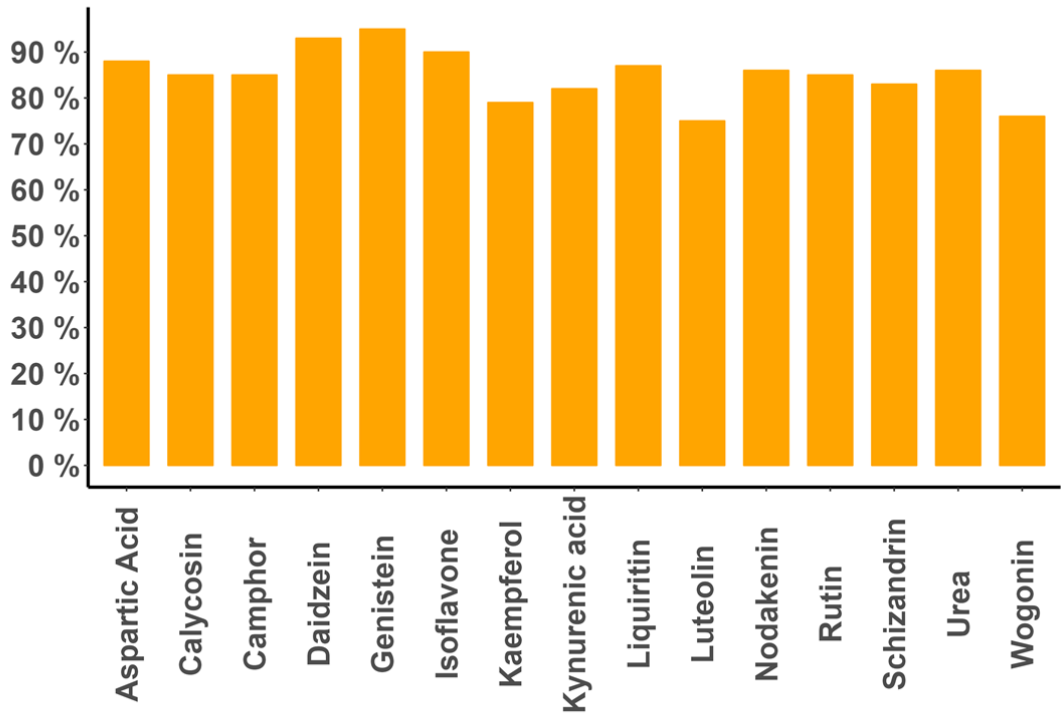

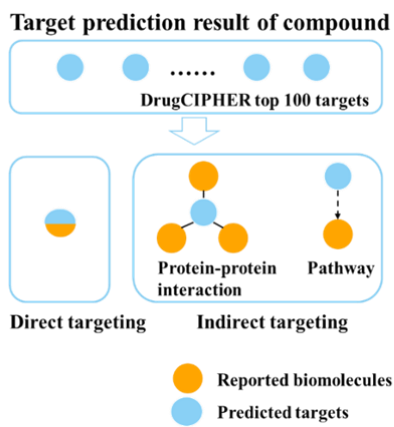

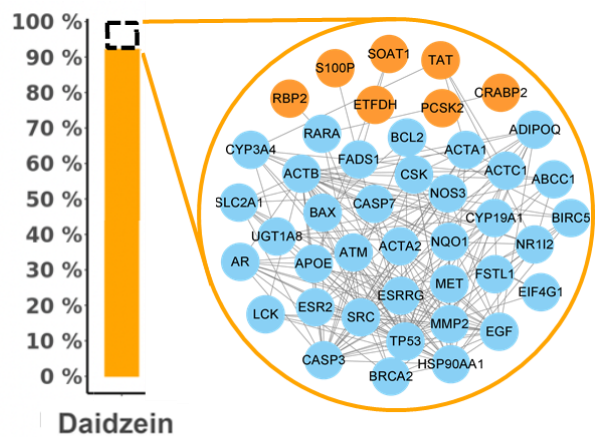

Figure 1. Targets prediction and literature validation of predicted targets of compounds in YQTQP. (A)Target prediction and validation of compounds in YQTQP, displaying literature coverage rate (76-95%) of the main chemical components. (B) The relationship between the predicted targets and the targets reported in the literature. (C) The reported and unreported predicted targets of Daidzein in the same biomolecular network

## 2.2 Network target analysis of YQTQP in the treatment of AR

Based on a statistic model proposed in previous study, the holistic targets of YQTQP was obtained to infer potential pathways and biological processes affected by YQTQP. Potential pathways and biological processes targeted by YQTQP were enriched through KEGG and GO enrichment. By integrating the results of enrichment analysis and prior knowledge of AR mined from related databases and literature, 12 potential targeted pathways and biological processes of YQTQP in the treatment of AR were identified and grouped in to three biological modules, including epithelial barrier dysfunction, cell adhesion, as well as regulation of immune and inflammation (Figure 2).

Among them, immune and inflammation were reported to be key links in AR[23], in which T helper (Th) 2 cells might be one of the vital immune cells and its infiltration might be main cause of late-phase allergic response[24, 25]. Apart from T cells, mast cells and their degranulation led by early-phase response was reported in previous studies[26]. Besides, biomolecules like IL-4, IFN-γ and TNF-α induced epithelial barrier disorder[27, 28] was important since intact skin and mucosal barriers are crucial for the maintenance of tissue homeostasis[29-31]. In other allergic diseases, epithelial barrier was also played vital role[32, 33]. And adhesion in cells promotes the differentiation as well as regulating nasal epithelial cells in AR[34-36]. In general, the results indicated that YQTQP potentially exerts therapeutic effects on AR through the pathways and biological processes involved in the activation, differentiation, and receptor signaling of immune cells, primarily Th1 and Th2 cells and mast cells, as well as the TNF signaling pathway-mediated inflammatory response in inflammation and immune-related events, which resulted in improved nasal immune environment and reduction of inflammation.

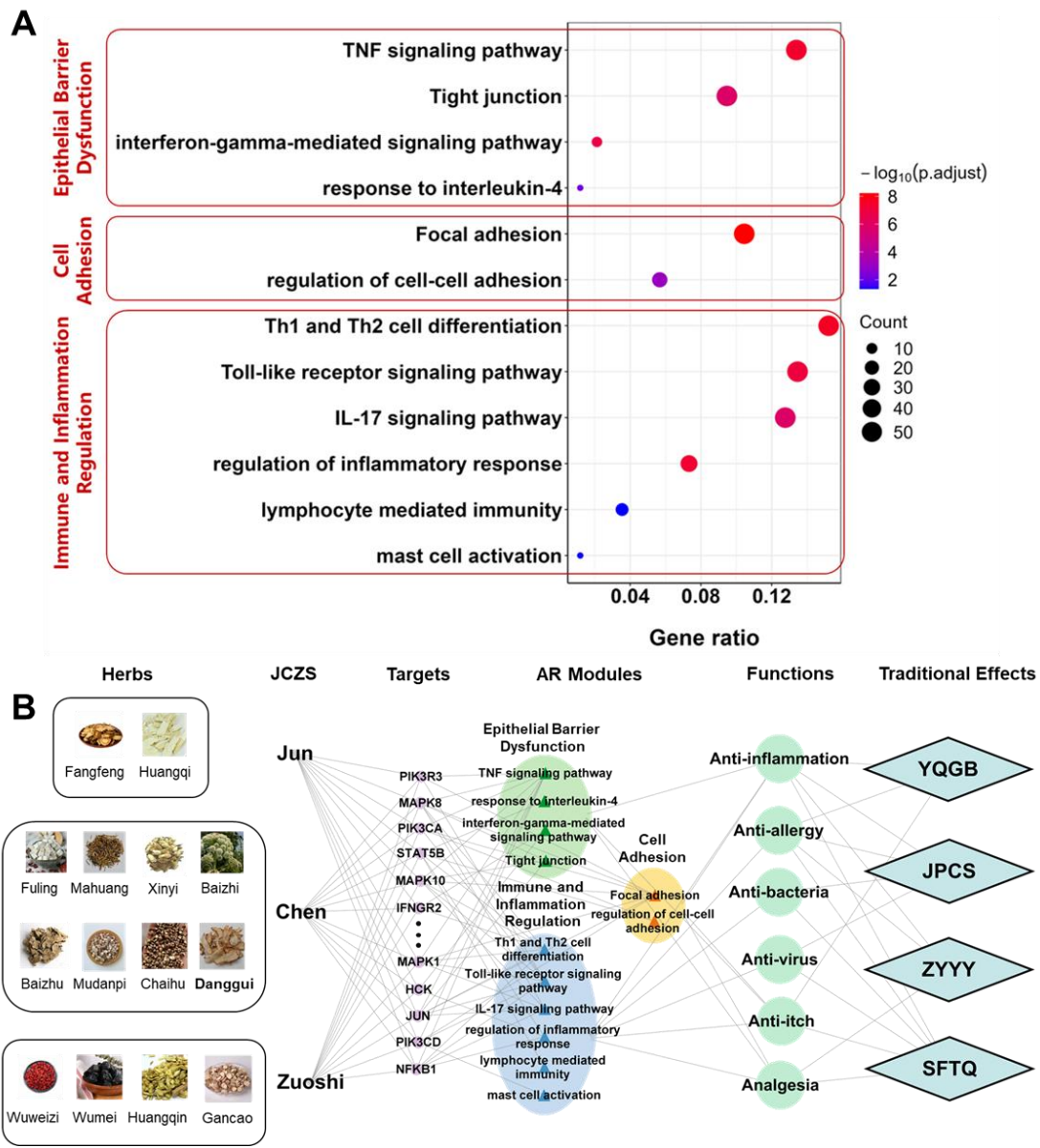

Figure 2. Network target analysis of YQTQP in modules of AR. (A) Enrichment result of pathways and biological processes in different modules of YQTQP in the treatment of AR. (B) Multi-layer network representing the Network target of YQTQP in the treatment of AR. YQGB, Yiqi Gubiao; JPCS, Jianpi Chushi; ZYYY, Ziyin Yangyang; SFTQ, Sanfeng Tongqiao.

According to the constructed multi-layer modular network, Network target of YQTQP in the treatment of AR was determined. During YQTQP's intervention of AR, through potential targets like PIK3s, MAPKs, IFNG and NFKB, YQTQP regulated AR from three aspects, including epithelial barrier dysfunction, cell adhesion, as well as immune and inflammation regulation. Thus, YQTQP played multiple roles in anti-inflammation, anti-allergy, anti-bacteria, anti-virus and analgesia. Besides, based on the combination of TCM theories, symptom genes collected from Gendoo[37] and information inferred from Symmap[31], relations between traditional therapeutic effects, modern effects and predicted modules were constructed. In this regard, we utilized the TCM theories to build the relationships among predicted modules, modern effects and traditional effects. For example, the effect of tonifying Qi (Yiqi Gubiao) and consolidating the exterior potentially corresponded to promoting the recovery of the body's epithelial barrier and preventing mucosal barrier damage, which is related to prevent bacteria and virus from invading the body; the effect of tonifying the spleen, dispelling dampness, nourishing Yin and Yang potentially corresponded to regulating the immune and inflammation of the body, improving the nasal immune environment, and reducing inflammation levels; and the effect of dispersing wind and opening the orifices potentially indicates that YQTQP may have a therapeutic effect on AR by improving cell adhesion, opening the orifices and feeling pain are always mentioned together in TCM theories. Additionally, the links between modules and functions are also validated in two ways. On one hand, we searched both the module names (or including pathway names) and functions in PubMed, and if the keys words of them appear in one abstract simultaneously, we regarded these two are related. On the other hand, enrichment analysis was performed on the symptom genes collected from Gendoo and Symmap, providing evidences for revealing the correlations between modules and functions in another way.

**2.3 Synergy pattern of herbs in YQTQP in the treatment of AR**

In addition to the analysis of the whole YQTQP, Network target analysis was also performed on each 14 TCM herbs in YQTQP based on the holistic target model. As shown in Figure 3, these herbs had potentially synergic effects on many pathways or biological processes, like TNF signaling pathway, tight junction, focal adhesion, Th cell differentiation, Toll-like receptor signaling pathway, IL-17 signaling pathway and regulation of inflammatory response. This indicated synergistic effects of multiple herbs in forming the effects of the whole formula which is consistent with Jun-Chen-Zuo-Shi (JCZS) principle of TCM. Besides, according to Network target analysis, Huangqi showed a dominant effect on all the three modules, while the other Jun herb, Fangfeng mainly play potential role in epithelial barrier dysfunction module and regulation of immune and inflammation module.

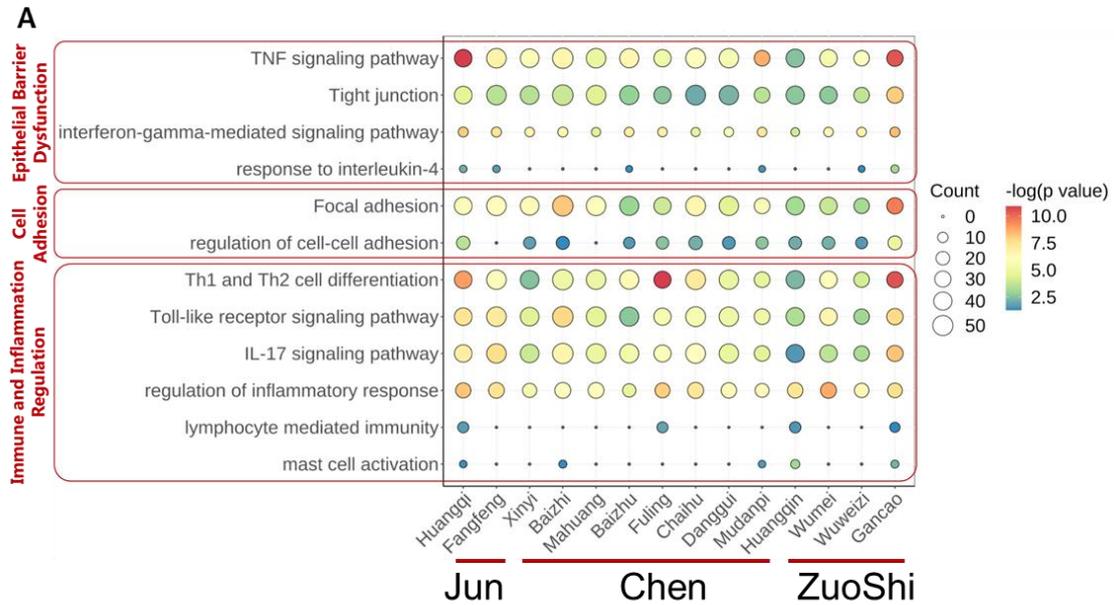

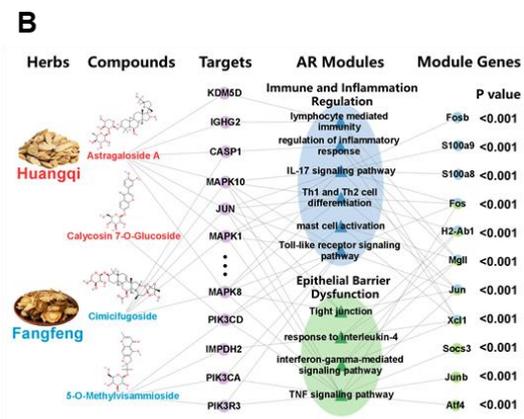

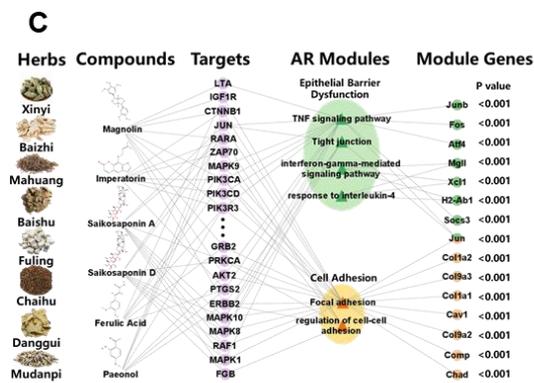

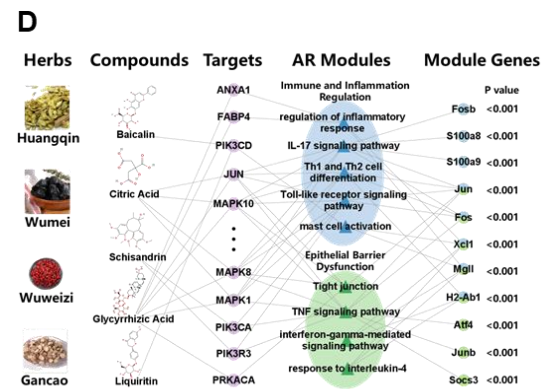

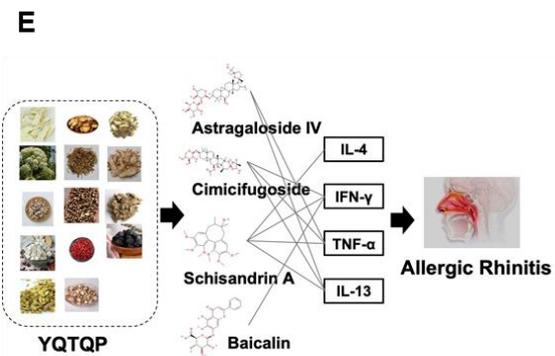

Figure 3. The mechanism of action of different JCZS and herbs of YQTQP in the treatment of AR. (A) Enrichment result of pathways and biological processes in different modules for different herbs in YQTQP in the treatment of AR. (B)-(D) Multi-layer network representing the Network target of JCZS of YQTQP in the treatment of AR, respectively. (E) Potential efficacy biomarkers and pharmacodynamic materials of YQTQP in the treatment of AR.

According to JCZS principle, herbal ingredients in a TCM formula are classified into 4 groups, in which Jun herbs are the main-acting herbs, whereas Chen, Zuo, and Shi herbs are herbs whose actions are to support those of Jun herbs in various ways such as enhancing therapeutic effects, reducing adverse effects, and delivering the effects to the target site. In YQTQP, it was found that representative compounds in Jun herbs, such as astragaloside IV and acteoside exerted potential effects on molecules such as CASP1, IGHG2, KDM5D, PI3K, and MAPKs, thereby exhibiting intervention on immune inflammation regulation and epithelial barrier disruption modules. Representative compounds in Chen herbs, such as magnoloside, schisandrol A, and ferulic acid, acted on targets such as LTA, IGF1R, CTNNB1, and PTGS2, thus affecting in key modules like epithelial barrier disruption and cell adhesion, whereas compounds in Zuo and Shi herbs, such as baicalin, mainly acted on molecules such as ANXA1, FABP4, and MAPKs, in the immune inflammation regulation and epithelial barrier disruption modules. A network connecting JCZS herbs, their key biological processes, and the targets through which they effect those processes, as well as corresponding traditional effects were illustrated in Figure 2B.

## 2.4 Inference of biomarkers of YQTQP in the treatment of AR

Following the network target analysis, the predicted targets and mechanism were verified through related literature and public dataset. Various studies have shown that IL-4[38-40], IFN-γ[41, 42], and TNF-α[43, 44] enhanced the mucosal permeability in both AR and other diseases, and pathways or biological processes related to these three modules have also been validated in the AR dataset GSE207084. Additionally, in

GSE207084, potential targets of YQTQP were found to have higher expression levels in AR epithelial cell samples. Key molecules in the three key modules of YQTQP, immune inflammation regulation, epithelial barrier disruption, and cell adhesion, were also found to have significant differential expression in the AR public dataset GSE207084 (Figure 4, $p < 0.001$). In a mouse model of house dust mite-induced allergic airway inflammation, antagonizing IL-4 can prevent mucosal barrier disruption and downregulation of tight junctions, thereby regulating biological processes related to epithelial barrier dysfunction [45]. Moreover, many components of the Jun herb, Huangqi, such as polysaccharides and astragalosides, have been found to have a regulatory effect on macrophages and lymphocytes [46].

The results of network target analysis showed that the key components in YQTQP and its corresponding TCM herbs were significantly enriched in the TNF, interleukin-related signaling pathways, and biological processes mediated by IFN-γ and IL-4. Disease databases, CTD and MalaCards, also indicated that IL-4, IFN-γ, TNF-α, and IL-13 were significantly associated with AR, which might be identified as potential pharmacodynamic biomarkers for YQTQP in the intervention of AR. Studies have shown that IL-4, IFN-γ, and TNF-α were associated with improved mucosal permeability in experimental mice. IL-4 and IL-13 were also common therapeutic targets in intervention of AR in clinical settings [28, 47, 48]. Based on the predicted pharmacodynamic effects of compounds in YQTQP as well as the enrichment results of the targets in the TNF signaling and Th2 cell differentiation pathways, together with biological processes mediated by IL-4 and IFN-γ, potential quality control materials for YQTQP in the intervention of AR was inferred with the guidance of Chinese Pharmacopoeia, including schisandrin B in Wuweizi, baicalin in Huangqin, Prim-O-glucosylcimifugin in Fangfeng, and astragaloside IV, were identified. The regulatory networks of the potential quality control materials and pharmacodynamic biomarkers are shown in Figure 4D.

## 2.4 Animal study

As mentioned above, key biomolecules like IL-4, IFN-γ, and TNF-α identified and verified to be key targets of YQTQP were highly associated with mucosal permeability in the pathogenesis of AR. Therefore, the animal study was conducted to further examine the effect of YQTQP in the treatment of AR. According to the relationships among modules of network targets, pharmacological effects and pathologic factors constructed in the multi-layer network, the anti-allergic and anti-inflammatory effects of YQTQP were measured in four aspects, including nasal mucosa allergy, abdominal vascular permeability, skin capillary permeability and granuloma (Figure 4A).

The analysis of tissue permeability in experimental mice revealed that YQTQP had significant intervention effects on inflammation, cell permeability, and allergic reactions. Specifically, at the inflammatory level, YQTQP lessened histamine-induced skin capillary permeability, and the effect increased with increasing dosage (Figure 4D). In terms of allergic reactions, the absorbance in the YQTQP group was significantly reduced, indicating a decrease in Evans blue extravasation, and the effect of YQTQP on allergic reactions was superior to that of the chlorpheniramine maleate group (Figure 4B). In terms of cell permeability, the high-dose group of YQTQP significantly reduced the permeability of peritoneal capillaries (Figure 4C, E). Compared with the saline group, high doses of this drug have a significant effect on reducing the weight of granulomas. This indicates that this drug can inhibit the proliferation of connective tissue during chronic inflammation (Figure 4F).

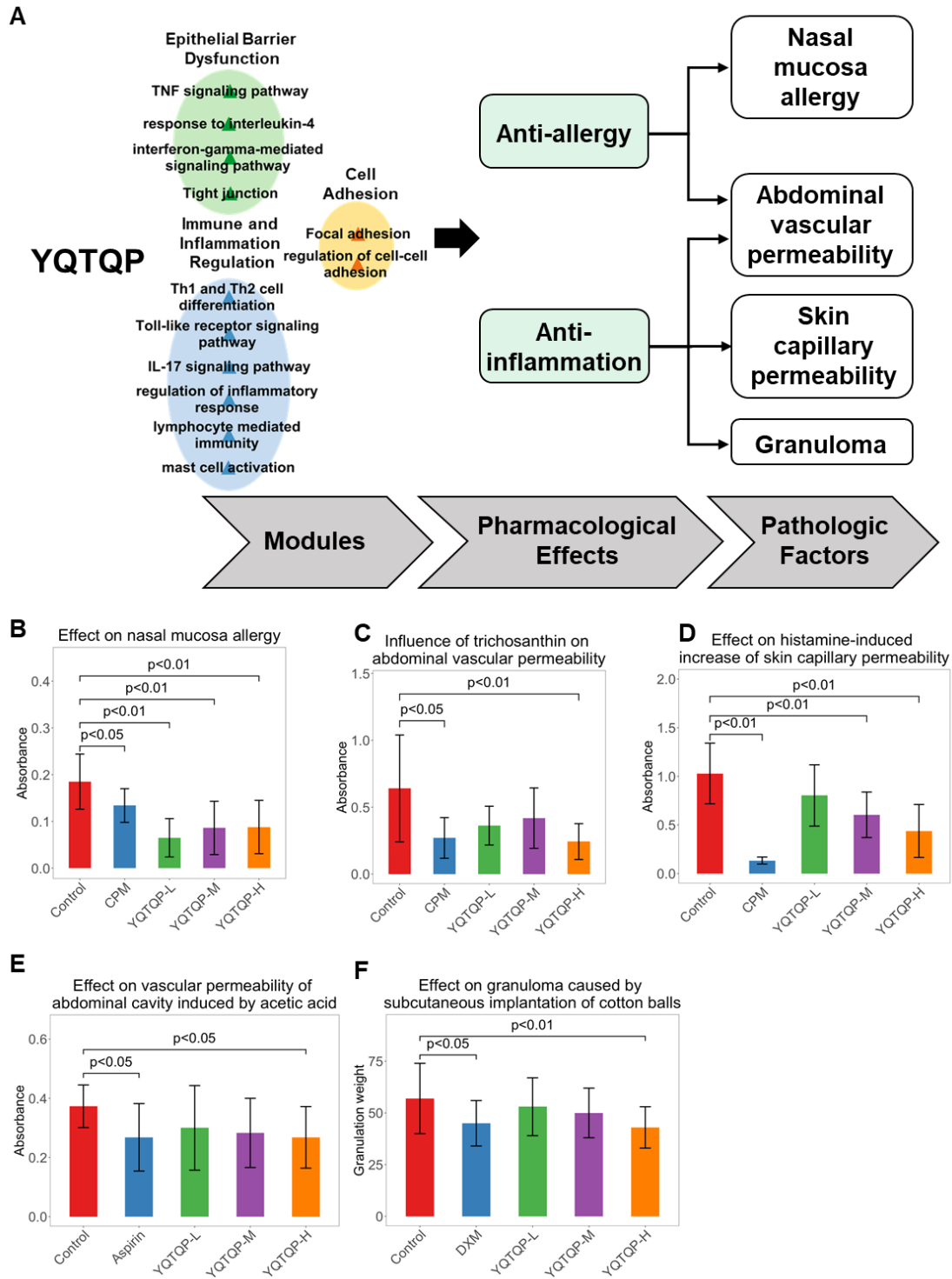

Figure 4. Verification of animal experimental results of YQTQP. (A) Hierarchical plot representing the relationships among modules of network targets of YQTQP, pharmacological effects and pathologic factors. (B)-(F) Interventional effects on different pathological indicators of YQTQP in the treatment of AR, including effect on nasal mucosa allergy, abdominal vascular permeability, skin capillary permeability, granuloma, itch.

## 2.5 Clinical study

From the perspective of TCM, the relationships between modules of network targets and four traditional efficacies were constructed based on the combination of TCM experience as well as enrichment analysis on the basis of Gendoo and Symmap. Besides, different clinical symptoms were measured for each traditional efficacy, like lack of strength, aversion to wind, tearing, abdominal distention, runny nose, etc. In order to validate these findings, clinical trials were carried out to validate the network targets from the macroscopic perspective.

A total of 480 participants were enrolled in this study (experimental group: 360, control group: 120), of which 458 completed the study (experimental group: 348, control group: 110), with 22 dropouts (experimental group: 12, control group: 10). The main reasons for dropout were loss of follow-up and poor efficacy. Except for one participant in the control group (subject 246) who violated the trial protocol by not using the experimental drug, all other cases were included in the full analysis set (FAS) and safety analysis set (SAS). The dropout rates for the experimental and control groups were 3.33% and 8.33%, respectively, and the difference between the groups was statistically significant but did not affect the evaluation of safety and efficacy between the two groups.

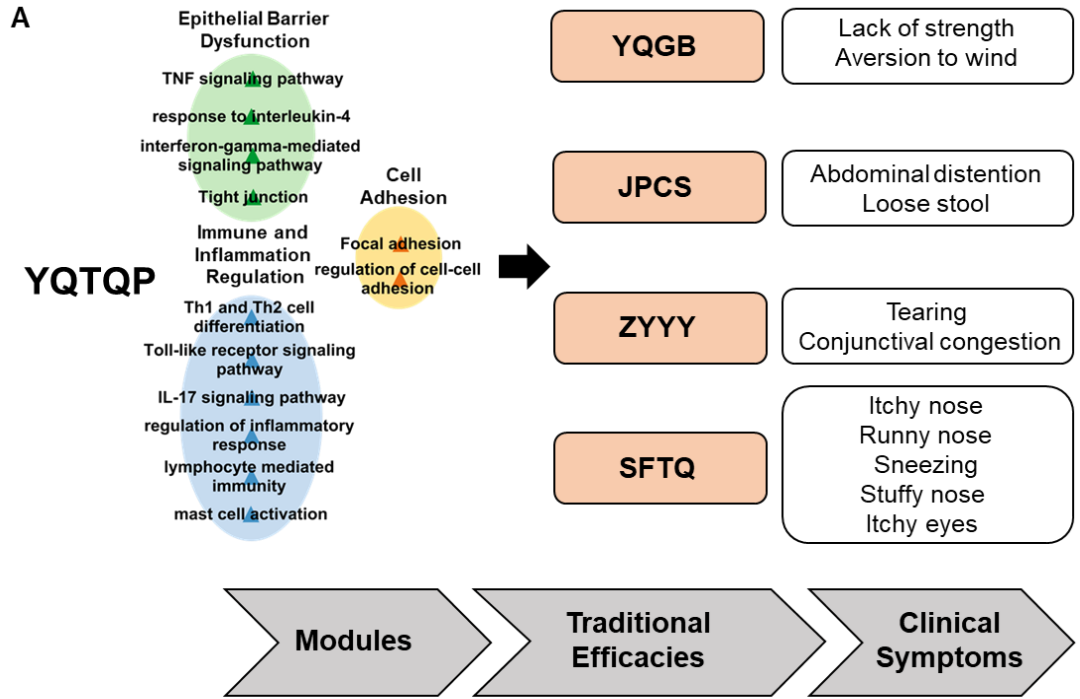

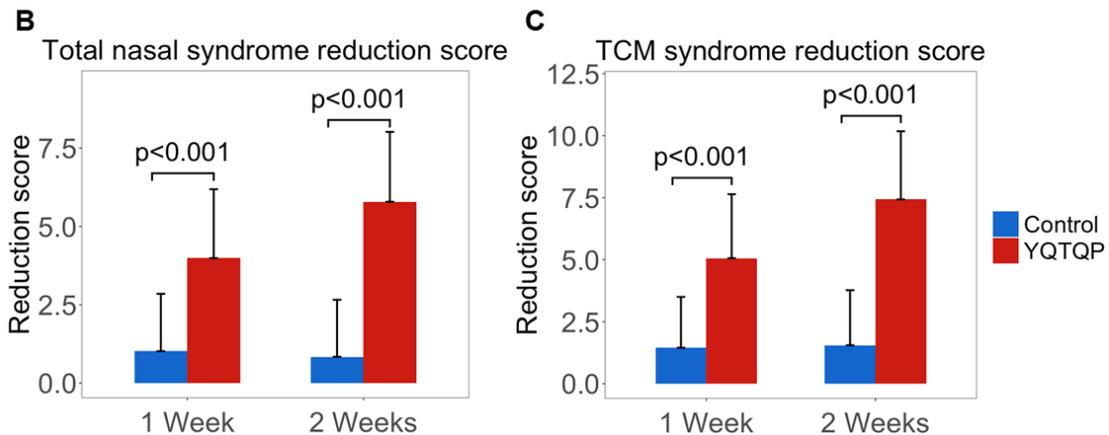

Figure 5. Verification of clinical efficacy test results of YQTQP. (A) Hierarchical plot representing the relationships among modules of network targets of YQTQP, traditional efficacy and clinical symptoms. (B) Bar plot of total nasal syndrome reduction score before and after 1week or 2 weeks of YQTQP intervention. (C) Bar plot of TCM syndrome reduction score before and after 1week or 2 weeks of YQTQP intervention.

In the FAS population, the total score of comprehensive nasal symptoms in the experimental group decreased by 3.99±2.20 after one week of treatment and 5.79±2.23 after two weeks of treatment, from the baseline score of 8.44±1.28. The corresponding values in the control group were 1.02±1.83 after one week of treatment and 0.84±1.82 after two weeks of treatment, from the baseline score of 8.24±1.28. There was a statistically significant difference between the two groups (P<0.0001). In the FAS population, the total score of TCM syndrome differentiation in the experimental group decreased by 5.06±2.58 after one week of treatment and 7.43±2.75 after two weeks of treatment, from the baseline score of 10.87±1.48. The corresponding values in the control group were 1.45±2.05 after one week of treatment and 1.55±2.22 after two weeks of treatment, from the baseline score of 10.82±1.56. There was a statistically significant difference between the two groups (P<0.0001). These results indicated that the effect of YQTQP in reducing the total score of comprehensive nasal symptoms and TCM syndrome differentiation was significantly better than that of the control group (Figure 5B, C). YQTQP can effectively alleviate seasonal AR and improve the related TCM syndrome of AR.

In detail, total nasal syndromes score, disappearance rates of four single indices were measured before and after the intervention of YQTQP, including rhinocnesmus (itchy nose), rhinorrhea (runny rose), sneeze and nasal obstruction (stuffy nose ). All these four indices were significantly decreased after the intervention of YQTQP (P<0.0001, Figure 6A-D), the same as local symptom scores for nasal examination (P<0.0001, Figure 6L). For eye accompanying symptoms, three indices inlcuing weep (tearing), itchy eyes, and conjunctival congestion were also significantly decreased (P<0.0001, Figure 6E-G). Besides, different syndromes in secondary syndrome of TCM before and after YQTQP intervention were also found to have higher disppearance rates after the intervention of YQTQP (P<0.0001, Figure 6H-K).

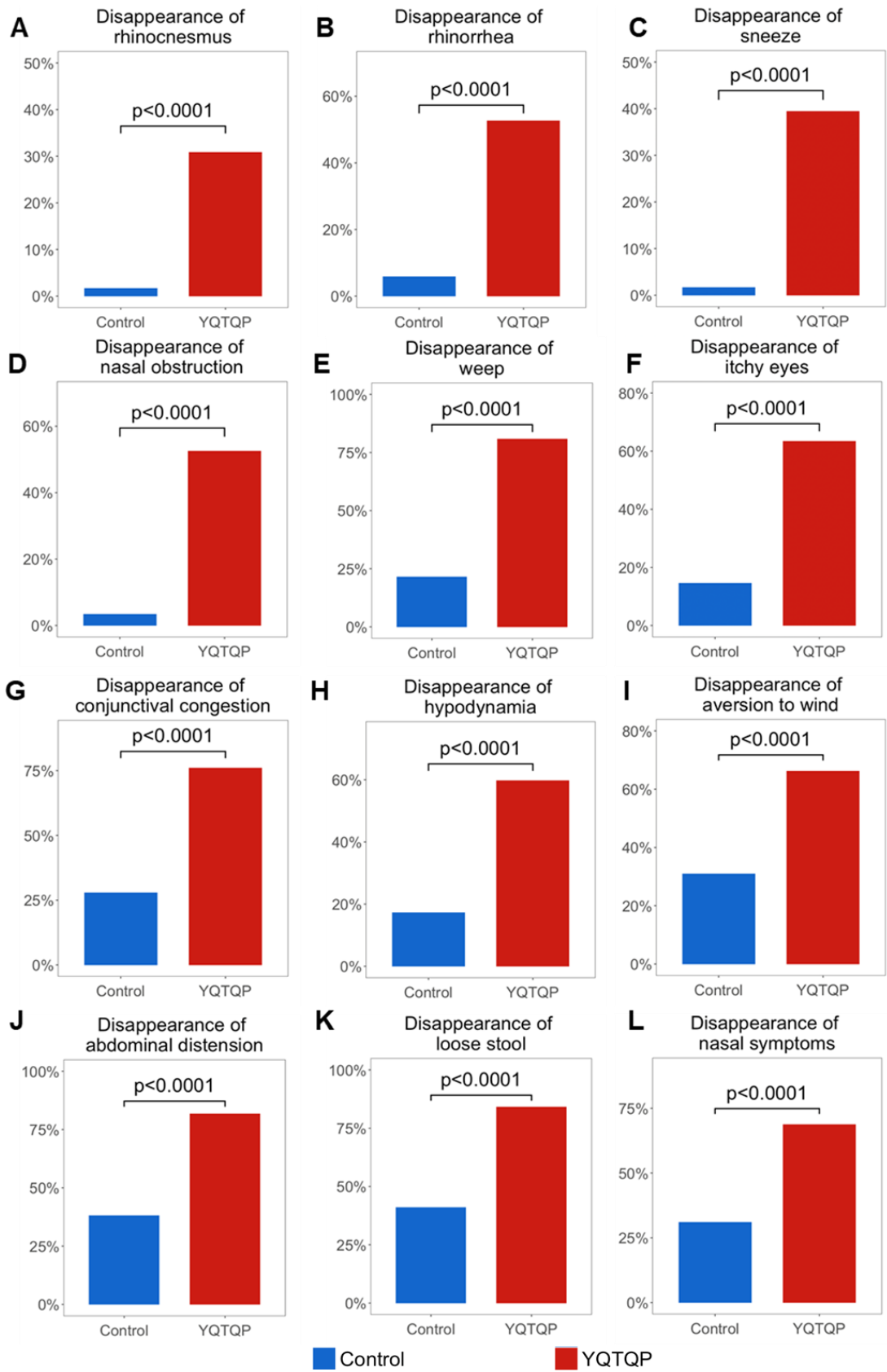

Figure 6. (A)-(D) Bar plots of disappearance rate of different syndromes in total nasal syndromes score before and after YQTQP intervention. (E)-(G) Bar plots of disappearance rate of different

syndromes in eyes before and after YQTQP intervention. (H)-(K) Bar plots of disappearance rate of different syndromes in secondary syndrome of TCM before and after YQTQP intervention. (L) Bar plots of disappearance rate of different syndromes in local symptom scores for nasal examination.

## Discussion

Regarding the unclear mechanism of action of YQTQP in the treatment of AR, this study established a network target model for YQTQP in the treatment of AR. The analysis revealed three key biological processes and pathways, namely, immune inflammation regulation, epithelial barrier dysfunction, and cell adhesion. Previous studies have shown that the nasal immune environment of AR is associated to biological processes such as immune cell activation, differentiation, receptor signal transduction mainly of Th1 and Th2 cells, mast cells, and inflammation mediated by the TNF signaling pathway. These key modules and molecules are common key target pathways and molecules for the treatment of AR[49], and also play an important role in other respiratory diseases [50] . Based on the network target model, the synergistic effects of different TCM herbs and Jun-Chen-Zuo-Shi roles of herbs in YQTQP on AR in the network target model was further revealed. Analysis showed that IL-4, IFN-γ, TNF-α, and IL-13 were pharmacodynamic markers for the intervention of YQTQP treating AR. Schisandrin A in *Schisandrae Chinensis Fructus*, baicalin in *Scutellariae Radix*, cimicifugoside in *Angelicae Sinensis Radix*, and astragaloside IV in *Astragali Radix* could serve as quality control markers for YQTQP.

The results of the animal study supported those of the prediction, which indicated that YQTQP had significant anti-allergic and anti-inflammatory effects, as well as certain counteractive effects against its complications such as allergic bronchospasm. Additionally, it showed auxiliary effects of analgesia and itching relief. Meanwhile, clinical study results have verified its efficacy in alleviating seasonal AR and improving the related TCM syndrome of AR.

In summary, Network target analysis suggested that YQTQP exerted pharmacological effects on AR through regulation of immune and inflammation process mediated by immune cells such as Th2 cells, mast cells, and other related pathways such as TNF signaling, IL-4, IFN-γ, as well as those related to epithelial barrier dysfunction and cell adhesion. Besides, this work might provide a new insight on revealing the mechanism of TCM based on the combination of multi-omics data, Network target analysis and experiments validation.

## Author contributions

S.L. and B.Z. supervised the study. S.L., B.Z. and D.F.H. designed the study. B.Y.W., D.F.Z. and T.Y.Z. performed the network target analysis. B.Y.W. performed the public omics datasets analysis. D.F.H. performed the animal experiments and clinical experiments. All authors discussed the results and wrote the manuscript.

## Declaration of Competing Interest

The authors declare that they have no known competing financial interests or personal relationships that could have appeared to influence the work reported in this paper.

## Acknowledgments

This work was supported by the National Natural Science Foundation of China, China [62061160369 and 81225025].